\documentclass[11pt]{article}
\usepackage{latexsym}
\begin{document}
\newcommand{\roughly}[1]%
       
\newcommand{\PSbox}[3]{\mbox{\rule{0in}{#3}
\includegraphics{#1}\hspace{#2}}}
\newcommand\lsim{\roughly{<}}
\newcommand\gsim{\roughly{>}}
\newcommand\CL{{\cal L}}
\newcommand\CO{{\cal O}}
\newcommand\half{\frac{1}{2}}
\newcommand\beq{\begin{eqnarray}}
\newcommand\eeq{\end{eqnarray}}
\newcommand\eqn[1]{\label{eq:#1}}
\newcommand\intg{\int\,\sqrt{-g}\,}
\newcommand\eq[1]{eq. (\ref{eq:#1})}
\newcommand\meN[1]{\langle N \vert #1 \vert N \rangle}
\newcommand\meNi[1]{\langle N_i \vert #1 \vert N_i \rangle}
\newcommand\mep[1]{\langle p \vert #1 \vert p \rangle}
\newcommand\men[1]{\langle n \vert #1 \vert n \rangle}
\newcommand\mea[1]{\langle A \vert #1 \vert A \rangle}
\newcommand\bi{\begin{itemize}}
\newcommand\ei{\end{itemize}}
\newcommand\be{\begin{equation}}
\newcommand\ee{\end{equation}}
\newcommand\bea{\begin{eqnarray}}
\newcommand\eea{\end{eqnarray}}

\def\Dsl{\,\raise.15ex \hbox{/}\mkern-12.8mu D}
\newcommand\Tr{{\rm Tr\,}}
\thispagestyle{empty}
\begin{titlepage}
\begin{flushright}
CALT-68-2400\\
\end{flushright}
\vspace{1.0cm}
\begin{center}
{\LARGE \bf  Diversification and Generalized Tracking Errors }\\ 
\bigskip
{\LARGE \bf For Correlated Non-Normal Returns}\\
~\\
\bigskip
\bigskip\bigskip
{ Mark B. Wise$^a$ and Vineer Bhansali$^b$} \\
~\\
\noindent
{\it\ignorespaces
          (a) California Institute of Technology, Pasadena CA 91125\\

          {\tt wise@theory.caltech.edu }\\
\bigskip   (b) PIMCO, 840 Newport Center Drive, Suite 300\\
               Newport Beach, CA 92660 \\

{\tt   bhansali@pimco.com}
}\bigskip
\end{center}
\vspace{1cm}
\begin{abstract}
The probability distribution for the relative return of a portfolio constructed from a subset $n$ of the assets from a benchmark, consisting of $N$ assets whose returns are multivariate normal, is completely characterized by its tracking error. However, if the benchmark asset returns are not multivariate normal then higher moments of the probability distribution for the portfolio's relative return are not related to its tracking error. We discuss the convergence of generalized tracking error measures as the size of the subset of benchmark assets increases. Assuming that the joint probability distribution for the returns of the assets is symmetric under their permutations we show that increasing $n$ makes these generalized tracking errors small (even though $n<<N$). For $n>>1$ the probability distribution for the portfolio's relative return is approximately symmetric and strongly peaked about the origin. The results of this paper generalize the conclusions of Dynkin et. al. (2002) to more general underlying asset distributions.
\end{abstract}
\vfill
\end{titlepage}
The excess return of portfolios managed against a benchmark can be attributed to active management in the selection of different weights from the benchmark, or from the selection of out-of-benchmark securities.  For a portfolio that attempts to replicate a benchmark's risk and performance with a subset of the benchmark securities under a portfolio replication approach, it is usually assumed that minimizing the tracking error of the portfolio with respect to the benchmark is equivalent to minimizing the underperformance or outperformance of the portfolio's relative return.  In principle, this replication approach suffers from the fundamental problem that the tracking error as a measure of risk relative to the benchmark fails in its essential purpose if the underlying asset returns are not joint normal, i.e. when it is needed the most.  In other words, if the asset returns are correlated and far from multivariate normal then it is possible that a tracking error optimized portfolio will not follow the benchmark as well as reasoning, based on the assumption that the portfolio's relative return is normal, would lead one to believe. But returns on corporate bonds subject to default risk or downgrade risk are not multivariate normal even if the underlying corporate assets are assumed to be since default of a company only occurs if the value of its corporate assets fluctuate below its liabilities\footnote{ This is usually described using a first passage time model. See, Black and Cox (1976), Longstaff and Schwartz (1995), Leland and Toft (1996), {\it etc}.}.  Furthermore, one does not have the luxury of increasing the number of assets in the portfolio ad infinitum since the addition of each new asset increases transactions costs and makes the portfolio less liquid.  Thus it is important to consider, for a wide range of asset probability distributions, how the total risk of a portfolio relative to the benchmark saturates as more assets from the benchmark are kept.

Consider a benchmark portfolio with $N$ assets that return $\hat r_i$, $i=1,\ldots N$. Assuming the value of the benchmark portfolio is equally divided amongst these assets the random variable for the benchmark portfolio return is
\be
\hat R={1 \over N}\sum_{i=1}^N\hat r_i.
\ee
Suppose a portfolio ${\cal P}$ consists of equal amounts of $n$ of these assets which, without loss of generality we choose to be $i=1,\ldots , n$. Then the random variable for the return of ${\cal P}$ relative to the benchmark is
\be
\Delta \hat R=\sum_{i=1}^N w_i \hat r_i,
\ee
where,
\be
w_i={1 \over n}-{1 \over N},
\ee
when $i=1,\ldots , n$ and
\be
w_i=-{1 \over N},
\ee
when $i=n+1,\ldots , N$. 
The joint probability distribution for the asset returns can be characterized by the expected values,
\be
r^{(m)}_{i_1 \ldots i_m}=E[\hat r_{i_1}\ldots \hat r_{i_m}].
\ee

We assume the joint probability distribution for the asset returns is symmetric under their permutations. This should make it as easy as possible for ${\cal P}$ to track the benchmark while still allowing for fluctuations in ${\cal P}$'s relative return. More realistic situations where the returns behave differently will not track the benchmark as well. With this assumption the expected asset returns are all equal, $r^{(1)}_i=r^{(1)}_1$, the diagonal elements of the asset return covariance matrix are all the same, $r^{(2)}_{ii}=r^{(2)}_{11}$, and the off-diagonal elements of the asset return covariance matrix\footnote{$r^{(2)}$ is related to the covariance matrix $\rho$ by, $\rho_{ij}=r^{(2)}_{ij}-r^{(1)}_i r^{(1)}_j$.}  are also independent of which asset is being considered, $r^{(2)}_{ij}=r^{(2)}_{12}$ for $i \ne j$. Note however we do not demand that the off-diagonal elements be equal to the diagonal ones since this would correspond to a correlation matrix that has one in all its elements. The generalization of this to higher values of $m$ is straightforward. Explicitly, $r^{(3)}_{iii}=r^{(3)}_{111}$, $r^{(3)}_{iij}=r^{(3)}_{112}$ for $i\ne j$ and $r^{(3)}_{ijk}=r^{(3)}_{123}$ for $i \ne j \ne k$, {\it etc}.

In the presence of correlations the central limit theorem cannot be used to argue that for $N>>1$ the benchmark return $\hat R$ becomes normally distributed or that for $n>>1$ ${\cal P}$'s return becomes normally distributed. It is sometimes argued that for large portfolios correlations should average out leaving one in a situation where effectively the central limit theorem is applicable. However, we consider that to be very unlikely. For example, corporate asset correlations, which determine the correlations of stock and corporate bond returns, are usually positive. This is partly for economic reasons but it also has an underlying mathematical origin. If a $N \times N $ correlation matrix has all its off-diagonal elements equal to $\xi$ then it is only mathematically consistent ({\it i.e}, has non-negative eigenvalues) for $- 1/(N-1) \le \xi \le 1$ [Wise and Bhansali (2002)]. In the limit $N \rightarrow \infty$ negative correlations are not allowed and hence the average correlation (found by averaging over allowed values of $\xi$ with any nonsingular probability distribution) will not be zero.

The expected value of ${\cal P}$'s relative return vanishes,
\be
E[\Delta \hat R]=r^{(1)}_1\left[n \left({1 \over n}-{1 \over N} \right)+(N-n)\left(- {1 \over N} \right) \right]=0.
\ee
However ${\cal P}$'s relative return does fluctuate. Define the moments of ${\cal P}$'s relative return probability distribution by,
\be
\label{sum}
v^{(m)}=E[(\Delta \hat R)^m]=\sum_{i_1,\ldots,i_m}w_{i_1}\ldots w_{i_m}r^{(m)}_{i_1 \ldots i_m}.
\ee
The second moment, $m=2$ is the square of the tracking error. For $m=2$ [Dynkin, Hyman and Konstantinovsky (2002)] ,
\begin{eqnarray}
\label{lehman}
v^{(2)}&=&r^{(2)}_{12}\left[ n(n-1) \left({1 \over n}-{1 \over N} \right)^2+2n(N-n)\left({1 \over n}-{1 \over N} \right) \left(-{1 \over N} \right) \right. \nonumber \\
&+&\left. (N-n)(N-n-1)\left(-{1 \over N}\right)^2                                                                              \right]+r^{(2)}_{11}\left[ n \left( {1 \over n}-{1 \over N} \right)^2+(N-n)\left(-{1 \over N } \right)^2 \right]\nonumber \\
&=&(r^{(2)}_{12}-r^{(2)}_{11})\left({1 \over N}-{1 \over n} \right).
\end{eqnarray}
For the coefficient of $r^{(2)}_{12}$ the first term in the square brackets comes from terms in the sum of equation (\ref{sum}) with both the subscripts of $r^{(2)}$ corresponding to assets in the portfolio ${\cal P}$, the second term comes from terms in the sum of equation (\ref{sum}) with one of the subscripts of $r^{(2)}$ corresponding to assets in the portfolio ${\cal P}$ and the third term comes from terms in the sum of equation (\ref{sum}) with none of the subscripts of $r^{(2)}$ corresponding to assets in the portfolio ${\cal P}$.

Taking the limit $N \rightarrow \infty$ gives,
\be
\label{simple}
v^{(2)}={1 \over n}(r^{(2)}_{11}-r^{(2)}_{12}).
\ee
Even with non-zero asset return correlations diversification achieved by increasing $n$ reduces the variance of ${\cal P}$'s return relative to the benchmark. This is very different from the behavior of the variance of ${\cal P}$'s return (without the benchmark subtracted) which, in the presence of significant correlations, cannot be made small by increasing $n$. 

We do not assume that the probability distribution for $\Delta R$ is normal and so the moments $v^{(m)}$ with $m>2$ are not related to the tracking error. The purpose of this paper is examine the behavior of these moments of the probability distribution for ${\cal P}$'s relative return when $N>>1$, $n >>1$ and $n/N<< 1$. This is particularly important in times of market stress for portfolios containing corporate bonds. In such eras default correlations are large [Das, Freed, Geng and Kapadia (2001)] and as remarked earlier the joint probability distribution of corporate bond returns is expected to be very far from multivariate normal.

We are interested in the value of the $m > 2$ moments of ${\cal P}$'s relative return as the benchmark portfolio gets arbitrarily large, $N \rightarrow \infty$, and write in that limit
\begin{eqnarray}
\label{series}
v^{(m)} &=& \alpha_0(m,n)~r^{(m)}_{12\ldots m}+\alpha_2(m,n)~r^{(m)}_{112\ldots m-1}+\alpha_3(m,n)~r^{(m)}_{1112\ldots m-2}\nonumber  \\
&+&\alpha_{2,2}(m,n)r^{(m)}_{11223\ldots m-2}+\ldots~~~.
\end{eqnarray}
The subscript on $\alpha$ denotes how many subscripts of $r^{(m)}$ take on the same value. For example $\alpha_{i_1,\ldots ,i_p}$ corresponds to $i_1$ indices taking on the same value (which without loss of generality can be taken to be $1$),  $i_2$ of the indices taking on the same value (which without loss of generality can be taken to be $2$), {\it etc}. Equation (\ref{simple}) yields, $\alpha_0(2,n)=-1/n$ and $\alpha_2(2,n)=1/n$. Performing a similar computation for the third moment yields, $\alpha_0(3,n)=2/n^2$, $\alpha_2(3,n)=-3/n^2$ and $\alpha_3(3,n)=1/n^2$.

Generalizing the $N \rightarrow \infty$ limit of the coefficient of $r^{(2)}_{12}$ in equation (\ref{lehman}) to arbitrary $m$ gives (assuming $n>m$),
\begin{eqnarray}
\alpha_0(m,n)&=&\sum_{k=0}^{m}\left({m! \over k!(m-k)!}\right)\left({n! \over (n-k)!}\right)(-1)^{m+k}\left({1 \over n}\right)^k \nonumber  \\
&=&(-1)^m n^{-m}U(-m,1-m+n,n), 
\end{eqnarray}
where U is the second solution to the confluent hypergeometric equation\footnote{See, for example, pages 753-758 of Arfken (1985).}. The $k=p$ term in the sum corresponds to terms in the sum of equation (\ref{sum}) where $p$ of the subscripts of $r^{(m)}_{i_1 \ldots i_m}$ label assets in the portfolio ${\cal P}$. The remaining $m-p$ subscripts label assets not in ${\cal P}$.
With $m$ held fixed and $n$ taken large $U(-m,1-m+n,n)$ is of order $n^{(m-1)/2}$ for $m$ odd and of order $n^{m/2}$ for $m$ even. 

This suggests that for $n>>1$,
\be
\label{mom}
(v^{(2)}) \sim O(1/n),~~{v^{(2m)}\over (v^{(2)})^m } \sim O(1){\rm ,~~ and~~} {v^{(2m+1)}\over (v^{(2)})^{(m+1/2)}} \sim O({1 \over {\sqrt n}}).
\ee
and it is not difficult to show this is indeed true by also considering the cases where some of the subscripts on $r^{(m)}_{i_1 \ldots i_m}$ take on the same value. If $j$ indices take on the same value then to get a contribution that survives in the limit $N \rightarrow \infty$ these indices must correspond to assets in ${\cal P}$ and so in this case,
\begin{eqnarray}
\alpha_j(m,n)&=&\left({m! \over j!(m-j)!}\right)\sum_{k=0}^{m-j}\left({(m-j)! \over k!(m-j-k)!}\right)\left({(n-1)! \over (n-1-k)!}\right)(-1)^{m-j+k}\left({1 \over n}\right)^{j-1+k} \nonumber  \\
&=&\left({m! \over j!(m-j)!}\right)(-1)^{m-j} n^{1-m}U(j-m,j-m+n,n). 
\end{eqnarray}
Since for large $n$ the function $U(j-m,j-m+n,n)$ is of order $n^{(m-j-1)/2}$ for $m-j$ odd and of order $n^{(m-j)/2}$ for $m-j$ even these terms leave unchanged the behavior in equation (\ref{mom}).  The other cases are handled similarly.

A simple example illustrates the power of the constraint on the probability distribution for ${\cal P}$'s relative return that equation (\ref{mom}) provides. Consider the probability distribution,
\begin{eqnarray}
\label{ex}
P(\Delta R)&=&{1 \over 2(1+ a/n)}\left[\delta \left(\Delta R-{1 \over 2 {\sqrt n}}\right)+\delta \left(\Delta R+{1 \over 2 {\sqrt n}}\right)+{a \over  n}\delta \left(\Delta R-{1 \over 2}\right)\right. \nonumber \\
&+&\left.{a \over  n}\delta \left(\Delta R+{1 \over 2}\right)\right],
\end{eqnarray}
where $\delta$ denotes the Dirac delta function and $a$ is a positive constant. For large $n$ the second moment of this probability distribution is $v^{(2)}\sim (1+a)/(4n)$ and for $m >1$ (and $n$ large) its $(2m)$'th moment is $v^{(2m)}\sim a/(2^{2m}n)$. While this probability distribution is consistent with the behavior the second moment in equation (\ref{mom}) for $a \ne 0$ it is in conflict with the behavior of the higher moments. With $a \ne 0$ the probability distribution in equation (\ref{ex}) implies that the probability for the relative return $\Delta R=\pm 1/2$ is of order $1/n$, which is too large to be consistent with the scaling of the higher moments in equation (\ref{mom}). 

We have generalized the results of Dynkin, Hyman and Konstantinovsky (2002), establishing that higher moments of ${\cal P}$'s relative return become successively smaller for $n>>1$ (by powers of $1/{\sqrt n}$) even if asset return correlations are large and the joint probability distribution for the asset returns is not multivariate normal. From equation (\ref{mom}) it is also evident that as $n$ gets large the skewness becomes small ({\it i.e.,} $O(1/{\sqrt n}$)). For  $N >>1$, $n >>1$ and $n/N<< 1$, the probability distribution for ${\cal P}$'s relative return is approximately symmetric and strongly peaked about zero. However, it does not necessarily approach a normal one ({\it e.g.} the excess kurtosis need not be small). For very large $n$ the scaling in equation (\ref{mom}) is consistent with the probability distribution for $\Delta R$ being a symmetric function of ${\sqrt n}\Delta R$.

\vskip0.25in
\newpage
\noindent
{\Large{\bf References}}
\vskip0.25in

\noindent
Arfken, G. (1985) {\it Mathematical Methods For Physicists}, Academic Press Inc.

\vspace{0.2cm}

\noindent
Black, F., and Cox, J. (1976) {\it Valuing Corporate Securities: Some Effects of Bond Indenture Provisions}, Journal of Finance, 31, 351-367.

\vspace{0.2cm}

\noindent
Das, S., Freed, L., Geng, G. and Kapadia, N. (2001) {\it Correlated Default Risk}, working paper, Department of Finance Santa Clara University and Gifford Fong Associates.

\vspace{0.2cm}

\noindent
Dynkin, L., Hyman, J. and Konstantinovsky, V. (2002) {\it Sufficient Diversification in Credit Portfolios}, Lehman Brothers Fixed Income Research.

\vspace{0.2cm}

\noindent
Leland, H. and Toft, K. (1996) {\it Optimal Capital Structure, Endogenous Bankruptcy and the Term Structure of Credit Spreads}, Journal of Finance, 51, 987-1019.

\vspace{0.2cm}

\noindent
Longstaff, F. and Schwartz, E. (1995) {\it A Simple Approach to Valuing Risky Floating Rate Debt}, Journal of Finance, 50, 789-819.

\vspace{0.2cm}

\noindent
Wise, M. and Bhansali, V. (2002) {\it Portfolio Allocation To Corporate Bonds with Correlated Defaults}, To appear in the Journal of Risk.

\vspace{0.2cm}

\end{document}